\documentclass[aps,superscriptaddress,showpacs,nofootinbib,twocolumn]{revtex4}
\usepackage{bm}
\usepackage{graphicx}

\begin{document}

\title{Mirror symmetry breaking and restoration: the role of noise and chiral bias}

\author{David Hochberg}
\email{hochbergd@inta.es}
\affiliation{Centro de
Astrobiolog\'{\i}a (CSIC-INTA), Carretera Ajalvir Kil\'{o}metro 4, 28850
Torrej\'{o}n de Ardoz, Madrid, Spain}

\begin{abstract}
The nonequilibrium effective potential is computed for the Frank
model of spontaneous mirror symmetry breaking (SMSB) in chemistry in
which external noise is introduced to account for random
environmental effects. When these fluctuations exceed a critical
magnitude, mirror symmetry is restored. The competition between
ambient noise and the chiral bias due to physical fields and
polarized radiation can be explored with this potential.
\end{abstract}

\date{\today}

\maketitle

Enantiomers are molecules that are nonsuperimposible complete mirror
images of each other. A remarkable feature of Nature is that
this mirror or chiral symmetry is broken in all biological
systems, where processes crucial for life such as replication, imply chiral
supramolecular structures, sharing the same
chiral sign (homochirality) for all present living systems. These chiral
structures are proteins, composed by aminoacids almost exclusively found as the left-handed
enantiomers (L), and DNA, RNA polymers and sugars with chiral building blocks composed
by right-handed (D) monocarbohydrates, and chiral amphiphiles forming membranes. This fact
has led to the widespread perception that the presence of handed or chiral molecules is a
unique signature of living systems. The emergence of this biological homochirality in
the chemical evolution from prebiotic to living systems is a tantalizing enigma in the origin of life, as is the
robustness of homochirality in actual living systems, and is a fascinating subject
that has intrigued scientists from diverse backgrounds.
Current reviews of the origin of homochirality can be found in
\cite{Franck,PCintas,Pod,Lahav,Bonner,Yus}.
Previous hypotheses suggesting that homochirality emerged after
the development of the primeval biological system \cite{Fox}, are being replaced
by the widespread conviction that enantiomerically pure compounds are a
prerequisite for the evolution of
living species and that mirror symmetry breaking must have taken place
before the emergence of life \cite{Bada,AGK,Orgel}.
We adopt the latter viewpoint here.

Frank introduced a paradigmatic model for spontaneous mirror
symmetry breaking (SMSB) and autocatalytic amplification in 1953
\cite{Frank}. A variant \cite{Gutman} of this open-flow reaction
scheme involves the two enantiomers L and D and an achiral reactant
A (kept at constant concentration) and the following reaction steps,
where the $k_{\pm i}$ denote the forward/reverse $(\pm)$ rate
constants: \textit{Production of chiral compound} $(k_1,k_{-1})$:
$\textrm{A} \leftrightarrow \textrm{L},\, \textrm{A} \leftrightarrow
\textrm{D}$, \textit{autocatalytic amplification} $(k_2,k_{-2})$:
$\textrm{L} + \textrm{A} \leftrightarrow \textrm{L} + \textrm{L},\,
\textrm{D} + \textrm{A} \leftrightarrow \textrm{D} + \textrm{D}$,
and \textit{mutual inhibition} $(k_3)$: $\textrm{L} + \textrm{D}
\rightarrow \textrm{LD}$. The heterodimer LD is removed from the
system. Frank's model contains the fundamental ingredients believed
to be essential for mirror symmetry breaking and subsequent chiral
amplification \cite{Blackmond}. It can be elaborated by adding in
polymerization side reactions \cite{Sandars,BM,WC,SH,Gleiser} that
can yield homochirality in populations of oligomers. The
corresponding rate equations expressed in terms of the enantiomeric
excess ($ee$) $\eta = ([L] - [D])/([L] + [D])$, the order parameter
for mirror symmetry breaking, and the net chiral matter $\chi = [L]
+ [D]$ are:
\begin{eqnarray}\label{eta}
d\eta/dt &=& - 2k_1A \eta/\chi + (k_3
-k_{-2})\chi\eta(1-\eta^2)/2 \\
\label{chi} d\chi/dt &=& 2k_1A - \chi^2[ k_{-2} +
(k_3-k_{-2})(1 - \eta^2)/2] \\
 &+& (k_2A - k_{-1})\chi. \nonumber
\end{eqnarray}
These rate equations are deterministic, but more realistic
treatments should take noise phenomena into account. The nature of
such fluctuations can be internal as well as external to the
chemical system. Intrinsic statistical fluctuations in $\eta$ about
the ideal racemic composition $[L]=[D]$ \cite{Mills}, as well as
diffusion-limited noise present in spatially extended systems
\cite{HZ1}, are sufficient to tip the system over into one of its
equally likely stable chiral states when $k_3 > k_{-2}$. For
prebiotic scenarios, the coupling of reaction schemes such as this
one to environmental effects (e.g., meteor impacts) is crucial for
determining the role of early planetary environments and external
disruptions on the emergence, if any, of homochirality
\cite{Gleiser}.

This Letter has a two-fold purpose. On the one hand, we aim to
establish analytically the impact of both environmental disturbances
and chiral bias on chemical systems that lead to SMSB. These
external effects can be modeled stochastically and lead one to
consider stochastic differential equations \cite{Gleiser}. Recently,
we developed an analytic perturbation method for calculating
potentials associated with a wide class of stochastic partial
differential equations \cite{HMPVa}. The potential is ideally suited
for treating symmetry breaking phenomena in nonequilibrium systems.
Hence, the second goal of this Letter is to demonstrate the
computational utility of that method for a fundamental model of
mirror symmetry breaking. The basic result is that ambient noise
tends to restore mirror symmetry and homochirality is diminished,
confirming independent numerical results \cite{Gleiser}.

\begin{figure}[hb]
\includegraphics[scale=0.68]{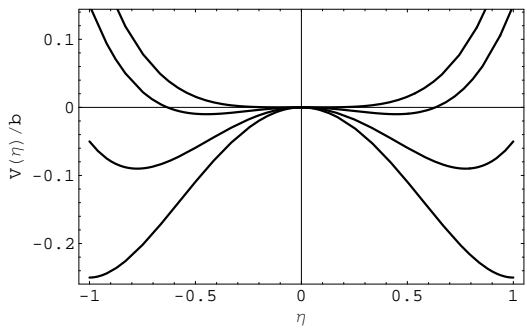}
\caption{\label{treepotfig} The Frank model potential $V(\eta)/b$,
Eq.(\ref{treepot}), displaying the racemizing tendency of the direct
chiral monomer production $\textrm{A} \rightarrow \textrm{L},\,
\textrm{A} \rightarrow \textrm{D}$. Sequence of curves from bottom
to top: $r=0, 0.2, 0.4$ and $r=0.5$.}
\end{figure}

\textit{Tree potential.} For constant $A$, introduce dimensionless
time $\tau = (k_2A - k_{-1})t$, and we verify that when the rate of
autocatalytic amplification exceeds the rate of monomer decay,
$\chi$ changes more rapidly than the enantiomeric excess $\eta$. The
system rapidly reaches a quasisteady state for $\chi$ $(d\chi/dt
\approx 0$) and then the slow variable $\eta$ evolves and the full
system reaches its true steady state \cite{HZ}. For this adiabatic
regime, we then put $\chi \rightarrow \chi^*$ in Eq. (\ref{eta}),
where $\chi^*$ denotes the quasisteady value for $\chi$. We define
the potential $V(\eta)$ \cite{BM} via $\frac{d \eta}{d t} = F(\eta)
= -V'(\eta)$, and so obtain
\begin{equation}\label{treepot}
\frac{V(\eta)}{b} = \frac{\eta^4}{4} +(r-\frac{1}{2})\eta^2 + v_0,
\end{equation}
where $v_0$ is an integration constant, and where $b =
\frac{1}{2}(k_3-k_{-2})\chi^* > 0$, $r = a/b$ and $a= k_1A/\chi^*
\geq 0 $. For the scaled potential, $r$ is the only free variable.
This is plotted in Fig. \ref{treepotfig} as a function of $-1 \leq
\eta \leq 1$  and for $0 \leq r \leq \frac{1}{2}$. The absolute
minima correspond to the asymptotic stable states of the chemical
system and are located at $\eta = \pm\sqrt{1-2r}$. By varying $r$,
we see how direct monomer production ($k_1 > 0$) tends to racemize
the system, as the two chiral minima move continuously towards zero
and coalesce at the origin when $k_1A$ increases. Strict
homochirality $|\eta| = 1$ holds only for $k_1A = 0$, otherwise,
$k_1A > 0$ implies $|\eta| < 1$.  For $r \geq \frac{1}{2}$, the
chiral symmetric state $\eta = 0$ is the only stable solution.
Gleiser and Walker \cite{Gleiser} obtained a potential qualitatively
similar to Fig. \ref{treepotfig}, for a reduced polymerization model
with direct production of monomers, which also clearly exhibits the
racemizing tendency of such autogenic terms (see their Fig. 1a).

\textit{The effective potential.} Following Gleiser and coworkers
\cite{Gleiser}, we couple the system to an external noise source
$\xi$ to model random environmental effects . Applying the methods
developed in \cite{HMPVa}, the corresponding stochastic differential
equation for $\eta$
\begin{equation}\label{stocODE}
\frac{d \eta}{d t} = F(\eta) + \xi(t), \qquad \langle
\xi(t)\xi(t')\rangle = {\cal A}\,\delta(t-t'),
\end{equation}
can be written as an ordinary differential equation with an
\textit{effective}, noise-corrected force $F_{\cal A}$, as follows :
\begin{equation}\label{effecODE}
\frac{d \eta}{d t} = F_{\cal A}(\eta),
\end{equation}
where to one-loop order in the noise amplitude $\cal A$, $F_{\cal
A}$ is given by \cite{HMPVa}
\begin{eqnarray}\label{efforce}
F_{\cal A}(\eta) &=& F(\eta) + \frac{1}{2}\frac{\cal A}{F(\eta)}
\Big( \Re \sqrt{[F'(\eta)]^2+F(\eta)F''(\eta)}\nonumber \\
&-& \sqrt{[F'(\eta)]^2} \Big) + O({\cal A}^2),
\end{eqnarray}
and $\Re$ is the real part. Thus for example, if a large meteor
impacts near a well-mixed prebiotic puddle or small pond, the bulk
pond is "shaken" as a whole and a time dependent noise $\xi(t)$
could provide a satisfactory description of the disturbance.

Eq.(\ref{treepot}) for $r=0$ implies $F(\eta) = b\eta(1-\eta^2)$.
The expression under the first square root in Eq.(\ref{efforce}) is
$[F'(\eta)]^2+F(\eta)F''(\eta) = b^2(1-12\eta^2+15\eta^4)$. This is
\textit{negative} on the open intervals $(-0.84, -0.31)$ and
$(0.31,0.84)$, zero on their endpoints, and is strictly positive
elsewhere \cite{footnote1}.

The one-loop effective potential is therefore given by
\begin{eqnarray}\label{Va}
V_{\cal A}(\eta) &=& -\int F_{\cal A}(\eta)\, d\eta + v_1,
\end{eqnarray}
where $v_1$ is an integration constant. We define
\begin{equation}
{\cal I}_1 = \int \frac{d\eta}{F(\eta)} \,
\sqrt{[F'(\eta)]^2+F(\eta)F''(\eta)}.
\end{equation}
This integral can be worked out in closed form and yields
\begin{eqnarray}\label{2I1}
2 {\cal I}_1 &=& -\ln|\frac{2\sqrt{R}-12\eta^2+2}{\eta^2}| + 2\ln
|\frac{4\sqrt{R}+18(\eta^2-1)+8}{\eta^2-1}| \nonumber \\
&-& \sqrt{15} \ln |2\sqrt{15R}+30\eta^2-12|,
\end{eqnarray}
valid whenever $R=1-12\eta^2+15\eta^4 \geq 0$. Otherwise, from
$\Re$ in Eq.(\ref{efforce}) we have ${\cal I}_1 = 0$.  Next, define ${\cal
I}_2$ as follows:
\begin{equation}
{\cal I}_2 = \int \frac{d\eta}{F(\eta)} \, \sqrt{[F'(\eta)]^2}.
\end{equation}
Since the function $\sqrt{[F'(\eta)]^2} = b|(1-3\eta^2)|$
then ${\cal I}_2$
\begin{eqnarray}\label{I2final}
 &=& \left\{ \begin{array}{c}
  \frac{1}{2}\ln|\eta^2| + \ln|1-\eta^2| + c_1,
   \,\, (-\frac{1}{\sqrt{3}} < \eta < \frac{1}{\sqrt{3}}) \\
    \\
   - \frac{1}{2}\ln|\eta^2| - \ln|1-\eta^2|+ c_2, \,\, (\eta \leq
-\frac{1}{\sqrt{3}}\, \& \, \frac{1}{\sqrt{3}} \leq \eta ).
 \end{array}
 \right.
\end{eqnarray}
Matching up at $\eta^2 = \frac{1}{3}$ ensures continuity in ${\cal
I}_2$. Without loss of generality, we take $c_2 = 0$. Then $c_1 =
-\ln \frac{1}{3} -2\ln \frac{2}{3} \simeq 1.91$.

The effective potential Eq.(\ref{Va}) can be written in terms of
these two integrals as follows:
\begin{equation}\label{effpot2}
V_{\cal A}(\eta) = V(\eta) -\frac{\cal A}{2}\left\{ {\cal I}_1
-{\cal I}_2 \right\} +  O({\cal A}^2),
\end{equation}
up to constants of integration used to match up the one-loop
corrections to insure continuity. For domains over which $R\geq 0$,
namely $\eta < -0.84$, and $-0.31 < \eta < 0.31$ and $\eta > 0.84$,
then ${\cal I}_1$ is given by Eq.(\ref{2I1}), otherwise when $R <
0$, then ${\cal I}_1 = 0$. Thus, for those regions over which $R<0$,
the one-loop correction in Eq.(\ref{effpot2}) is equal to
$+\frac{\cal A}{2} {\cal I}_2$. On the two outer intervals $(-1,
-0.84)$ and $(0.84, 1)$, the one-loop correction is given by
$-\frac{\cal A}{2}\left\{ {\cal I}_1 - {\cal I}_2 \right\}$. From
Eqs.(\ref{2I1},\ref{I2final}) we can calculate this quantity valid
on these intervals, and find that
\begin{eqnarray}\label{out}
\left\{ {\cal I}_1 -{\cal I}_2 \right\} &=&
-\frac{1}{2}\ln|\frac{2\sqrt{R}-12\eta^2+2}{\eta^2}| +
\frac{1}{2}\ln|\eta^2|\nonumber \\
&+& \ln \mid
4\sqrt{R}+18(\eta^2-1)+8\mid \nonumber \\
&-& \frac{\sqrt{15}}{2} \ln |2\sqrt{15R}+30\eta^2-12| .
\end{eqnarray}
Whereas for the central interval $(-0.31, 0.31)$, we calculate
\begin{eqnarray}\label{in}
\left\{ {\cal I}_1 -{\cal I}_2 \right\} &=&
-\frac{1}{2}\ln|2\sqrt{R}-12\eta^2+ 2| - \ln|1-
\eta^2|-c_1\nonumber \\
&+&  \ln \mid\frac{
4\sqrt{R}+18(\eta^2-1)+8}{\eta^2-1}\mid \nonumber \\
&-& \frac{\sqrt{15}}{2} \ln |2\sqrt{15R}+30\eta^2-12|.
\end{eqnarray}
Next write $V_{\cal A} = V + ({\cal A}/2) \Delta V$, then the form
of the pure one-loop correction $\Delta V(\eta)$ is completely
specified as follows:
\begin{eqnarray}\label{pureoneloop}
\Delta V(\eta)= \left\{
   \begin{array}{cc}
     \delta V_{out}(\eta) + v_1: & (-1, -0.84)  \\
     {\cal I}_2(\eta) + v_2:& (-0.84,-0.31) \\
     \delta V_{in}(\eta) + v_3: & (-0.31, 0.31) \nonumber \\
     {\cal I}_2(\eta) + v_2:& (0.31, 0.84) \\
     \delta V_{out}(\eta) + v_1: & (0.84, 1). \\
   \end{array}
 \right.
\end{eqnarray}
Here, $-\delta V_{out}$ is given by Eq.(\ref{out}), $-\delta V_{in}$
by Eq.(\ref{in}) and ${\cal I}_2$ by Eq.(\ref{I2final}). Matching up
at the endpoints of the above intervals fixes the constants $v_2 =
v_1+3.182$, $v_3 = v_1 -0.001$, where $v_1$ is an overall integration
constant we are free to choose; see Eq.(\ref{Va}). We take
$v_1 = \delta V_{in}(0)$.

\begin{figure}[h]
\includegraphics[scale=0.68]{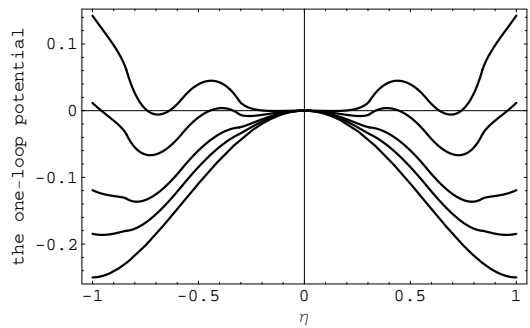}
\caption{\label{onelooppot} The one-loop effective potential
$V_{\cal A}/b$ for the Frank model. The final enantiomeric excess
$\eta$, corresponds to the absolute minima of the potential, and
decreases in absolute value below unity as the noise strength
increases. The curves from bottom to top correspond to ${\cal
A}/{2b} = 0.0, 0.05,0.1, 0.2$ and $0.3$.}
\end{figure}

\textit{Racemization.} We investigate the role that weak external
noise has on mirror symmetry breaking using the effective potential.
We first scale out by the factor $b$, and evaluate $V_{\cal A}/b$
while varying the dimensionless noise amplitude $0 \leq \frac{\cal
A}{2b} \ll 1$.  The absolute minima of the effective potential
correspond to the possible stable final chemical states. From the
sequence of curves in Fig. \ref{onelooppot}, corresponding to
$\frac{\cal A}{2b} = 0.0, 0.05, 0.1, 0.2$ and $0.3$, we see that
increasing the noise amplitude tends to racemize the system. The
homochiral states $|\eta| = 1$ exist only in the absence of noise
(bottom curve). For low levels of noise, the system has stable
chiral states corresponding to $|\eta| < 1$. For noise above a
critical value, the only stable final state is the racemic solution
$\eta = 0$ (top curve). Applying a linear stability analysis to
Eq.(\ref{effecODE})
we calculate the critical noise level ${\cal A}_c$ above which the
\textit{racemic state is the globally stable solution}: $\frac{{\cal
A}_c}{2b} = \frac{1}{3} \cong 0.33$, which is borne out by
inspection of the curves in Fig. \ref{onelooppot}. New relative
maxima begin to form and persist for ${\cal A}/2b \gtrsim {\cal
A}_c/2b$  thus leading to a pair of metastable chiral states (Fig.
\ref{onelooppot}). Numerical simulations, in two and three
dimensions \cite{Gleiser}, indicate however that the $ee$ goes to
zero continuously as the noise increases from zero and becomes
strong, so these extra maxima are most likely artifacts of the
lowest order calculation. Using the nominal values $k_3 \approx
10^{2} \,{\rm M}\, {\rm s}^{-1}$, $k_{-2} \approx 10^{-5} \,{\rm M}
\,{\rm s}^{-1}$ and $\chi^* = 1\, {\rm M}$, then $2b = 100 {\rm
s}^{-1}$, and external noises with ${\cal A} \lesssim 33 {\rm
s}^{-1}$ would be perturbatively valid.

\begin{figure}[h]
\includegraphics[scale=0.68]{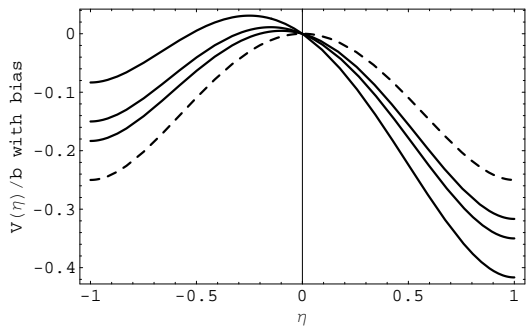}
\caption{\label{biaspoten} Tree potential subject to small chiral
bias. Dashed curve corresponds to zero bias. For $\epsilon'>0$, the
point $\eta=1$ is the absolute minimum of the potential. For the
sequence of solid curves below dashed curve, from top to bottom,
$\epsilon' = 0.1, 0.2$ and $0.5$. }
\end{figure}

\textit{Chiral bias.} External magnetic, electric, gravitational
fields, and vortex motion, as well as polarized radiation, can
induce mirror symmetry breaking \cite{Avalosb}. Chiral bias can be
studied via the potential by assigning chiral specific reaction
rates to the monomer production and autocatalysis steps thus
replacing $k_{i}$ by $k_i^{L} = k_i(1+\frac{1}{2}\epsilon)$ and
$k_i^{D} = k_i(1-\frac{1}{2}\epsilon)$ for $i = \pm 1, \pm 2$ where
$k_i=(k_i^L + k_i^D)/2$ \cite{KondeNelson,Gleiser}. For example,
$\epsilon = \frac{\Delta E}{kT} \approx 10^{-17}$ for parity
violation in the electroweak interactions at room temperature, where
$\Delta E$ is the energy difference between the two enantiomers
\cite{KondeNelson}. In the presence of chiral bias, the tree-level
potential is given by (for $r=0$) $\frac{V(\eta)}{b} =
\frac{\eta^4}{4}-\frac{\eta^2}{2} -\epsilon'[\eta -
\frac{\eta^3}{3}] + v_0$, where $\epsilon' =
\epsilon\frac{(k_2A-\frac{1}{2}k_{-2}\chi^*)}{(k_3-k_{-2})\chi^*}.$
This is plotted in Fig. \ref{biaspoten} for $1 > \epsilon' \geq 0$:
Due to the tilt, there are no longer racemic solutions for any bias
$\epsilon > 0$, only chiral states are possible, and only one of
these two chiral states will be an absolute minimum; see also Fig.
1b of Gleiser and Walker \cite{Gleiser}.
When we include noise, the one-loop biased effective potential is
obtained by subtracting $\epsilon'(\eta -\eta^3/3)$ from the right
hand side of Eq.(\ref{effpot2}), and is valid up to terms of order
$O(\epsilon' {\cal A})$ and $O({\cal A}^2)$. At this lowest order,
the effect of the bias is to tilt the noise corrected potential in
the same sense as shown in Fig. \ref{biaspoten}, so that the
sequence of noise induced minima located at $0 < \eta \leq 1$ in
Fig. \ref{onelooppot} now become the \textit{absolute} minima. Due
to this tilting the origin of the potential is no longer locally
flat ($V'_{\cal A}(0) \neq 0$) for any value of the noise. The noise
has a racemizing effect upon the biased system such that above a
critical noise level, the effective potential possesses a global
minimum corresponding to a weakly chiral state. Thus for example, we
calculate that for $\epsilon' = 0.001, 0.01$ and $0.1$ then
$\frac{{\cal A}_c}{2b} = 0.311,0.318$ and $0.388$, and the
corresponding enantiomeric excesses are $\eta \approx 0.10, 0.15$
and $0.20$, respectively. For increasing chiral bias, ever stronger
noise levels are ``tolerated" before homochirality $|\eta| = 1$ is
erased. A detailed account of noise and chiral bias on mirror
symmetry breaking will be provided elsewhere.

In this Letter we applied the stochastic field theory formalism of
\cite{HMPVa} to study the emergence of chirality in a key model of
SMSB in chemistry in which environmental effects are modeled by
external noise. We focused on the Frank model due to the central
role it plays in theoretical approaches to mirror symmetry breaking
\cite{AGK,Frank,Blackmond,Sandars,BM,
WC,SH,Gleiser,HZ1,HZ,Avalosb,KondeNelson}. By strictly analytic
means we verified that weak noise racemizes the system, erasing
homochirality. This is a perturbatively valid key result, confirming
the previous numerical results obtained by Gleiser and coworkers
\cite{Gleiser}. We also studied the competition between chiral bias
and external noise and verified that stronger noise levels are
required to racemize the system in the presence of bias. We assumed
well-mixed conditions (zero dimensional systems), but the analytic
method \cite{HMPVa} enjoys the flexibility to include diffusion in
$d$-dimensions and spatially dependent noise terms. A preliminary
study of the $d=2$ potential indicates that the results presented
here carry over when spatial dependence is included. The important
role that fluctuation phenomena, noise and chiral bias play in the
origin of homochirality can therefore be analyzed in an elegant and
systematic way.

We thank Josep M. Rib\'{o} and Mar\'{i}a-Paz Zorzano for useful
discussions, and acknowledge the Grant AYA2006-15648-C02-02 from the
Ministerio de Ciencia e Innovaci\'{o}n (Spain).

\end{document}